\documentclass[floatfix,a4paper,prl,twocolumn,superscriptaddress, showpacs,preprintnumbers,amsmath,amssymb]{revtex4}

\usepackage{hyperref}
\usepackage{amsmath}
\usepackage{amssymb}
\usepackage{graphicx, graphics}
\usepackage{color}
\usepackage{multirow}
\usepackage{dsfont}
\usepackage{mathbbol}

\begin{document}
\preprint{}

\title{Magnetic field tuning and quantum interference in a Cooper pair splitter}

\author{G.~F\"{u}l\"{o}p}
\thanks{These authors contributed equally to this work.}
\affiliation{Department of Physics, Budapest University of Technology and Economics, and Condensed Matter Research Group of the Hungarian Academy of Sciences, Budafoki \'{u}t 8, 1111 Budapest, Hungary}

\author{F. Dom\'inguez}
\thanks{These authors contributed equally to this work.}
\affiliation{Departamento de F\'isica Te\'orica de la Materia Condensada,
        Condensed Matter Physics Center (IFIMAC), \\ and Instituto Nicol\'as
        Cabrera, Universidad Aut\'onoma de Madrid, E-28049 Madrid, Spain}
    
\author{S.~d'Hollosy}
\thanks{These authors contributed equally to this work.}
\affiliation{Department of Physics, University of Basel, Klingelbergstrasse 82, CH-4056 Basel, Switzerland}
        
\author{A.~Baumgartner}
\email{andreas.baumgartner@unibas.ch}
\affiliation{Department of Physics, University of Basel, Klingelbergstrasse 82, CH-4056 Basel, Switzerland}

\author{P. Makk}
\affiliation{Department of Physics, Budapest University of Technology and Economics, and Condensed Matter Research Group of the Hungarian Academy of Sciences, Budafoki \'{u}t 8, 1111 Budapest, Hungary}
\affiliation{Department of Physics, University of Basel, Klingelbergstrasse 82, CH-4056 Basel, Switzerland}

\author{M.H. Madsen}
\altaffiliation[present address: ]{Danish Fundamental Metrology, DK-2800 Kgs. Lyngby, Denmark}
\affiliation{Center for Quantum Devices \& Nano-Science Center, Niels Bohr Institute, University of Copenhagen, Universitetsparken 5, DK-2100
Copenhagen, Denmark} 
        
\author{V.A.~Guzenko}
\affiliation{Laboratory for Micro- and Nanotechnology, Paul Scherrer Institute, CH-5232 Villigen PSI, Switzerland}

\author{J.~Nyg{\aa}rd}
\affiliation{Center for Quantum Devices \& Nano-Science Center, Niels Bohr Institute, University of Copenhagen, Universitetsparken 5, DK-2100
Copenhagen, Denmark}

\author{C.~Sch\"{o}nenberger}
\affiliation{Department of Physics, University of Basel, Klingelbergstrasse 82, CH-4056 Basel, Switzerland}

\author{A. Levy Yeyati}
\affiliation{Departamento de F\'isica Te\'orica de la Materia Condensada,
        Condensed Matter Physics Center (IFIMAC), \\ and Instituto Nicol\'as
        Cabrera, Universidad Aut\'onoma de Madrid, E-28049 Madrid, Spain}
        
\author{S.~Csonka}
\affiliation{Department of Physics, Budapest University of Technology and Economics, and Condensed Matter Research Group of the Hungarian Academy of Sciences, Budafoki \'{u}t 8, 1111 Budapest, Hungary}

\date{\today}

\begin{abstract}
Cooper pair splitting (CPS) is a process in which the electrons of naturally occurring spin-singlet pairs in a superconductor are spatially separated using two
quantum dots. Here we investigate the evolution of the conductance correlations 
in an InAs CPS device in the presence 
of an external magnetic field. In our experiments the gate dependence of the signal that depends on both quantum dots continuously evolves from a slightly asymmetric Lorentzian to a strongly asymmetric Fano-type
resonance with increasing field. 
These experiments can be understood in a simple three - site model, which shows that the nonlocal CPS leads to symmetric line shapes, while the local transport processes can exhibit an asymmetric shape due to quantum interference. These
findings demonstrate that the electrons from a Cooper pair splitter can propagate coherently after their emission from the superconductor and how a magnetic field can be used to optimize the performance of a CPS device. In addition, the model calculations suggest that the estimate of the CPS efficiency in the experiments is a lower bound for the actual efficiency.
\end{abstract}


\maketitle

In the Cooper pair splitting (CPS) process the electrons of the Cooper pairs in a superconductor are separated spatially using two
quantum dots (QDs) coupled in parallel to a central 
superconductor contact (S) in a three-terminal geometry
\cite{Recher_Loss_PRB63_2001, lesovik_martin_blatter_EPJB01, Sauret_2004_PRB70_2004}, 
see Fig.~1a. The Coulomb repulsion 
on the QDs and the quasiparticle energy gap of the superconductor
enforce the electrons to separate into different normal metal electrodes (N1 and N2). 
Since Cooper pairs are spin singlet states, 
such devices could serve as a source of nonlocal spin entangled 
electron pairs. Similar geometries are also relevant in the search 
for Majorana bound states \cite{Mourik_Kouwenhoven_Science_2012}, 
both, in a local S-N junction experiment or in a three-terminal 
setup, where an increase in CPS efficiency might serve as a signature of the elusive 
exotic states \cite{Nilsson_Akhmerov_Beenakker_PRL101_2008}.

In a series of recent experiments on semiconducting nanowires (NWs)
\cite{Hofstetter2009, Hofstetter_Baumgartner_PRL107_2011, 
Das_Heiblum_NatureComm_2012, Fulop_dHollosy_Baumgartner_PRB90_2014}, 
carbon nanotubes \cite{Herrmann_Kontos_Strunk_PRL104_2010, Schindele_Baumgartner_Schoenenberger_PRL109_2012, Schindele_Baumgartner_PRB89_2014}, 
and graphene \cite{Tan_Hakonen_arXiv_2014}, 
CPS was demonstrated by positive conductance correlations between the currents from S into the two N-contacts. In these experiments, external magnetic fields 
were solely used to suppress the superconductivity for control 
experiments, but not as a parameter to tune CPS. In addition, most experiments 
were interpreted in terms of an incoherent picture with independent 
transport mechanisms, only coupled by the quantum dot (QD) dynamics 
\cite{Schindele_Baumgartner_Schoenenberger_PRL109_2012, Fulop_dHollosy_Baumgartner_PRB90_2014}.

\begin{figure}[b]
\centering
\includegraphics[width=3.in,clip]{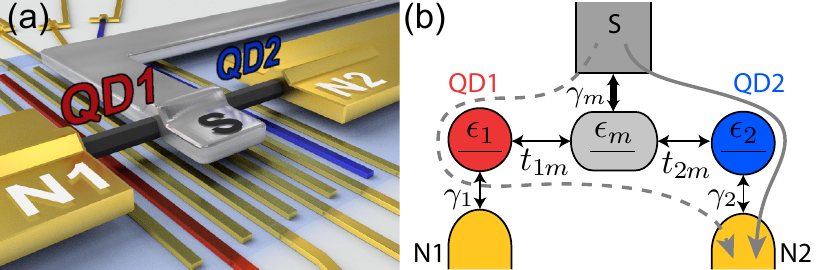}
\caption{(Color online)  (a) and (b) Schematics of 
a Cooper pair splitter and the corresponding 3-site 
model. The dashed and solid gray arrows represent two different interfering single-electron 
paths from S to N2, which can result in Fano-type conductance resonances.}
\label{Figure1}
\end{figure}

Here we report experiments in a NW-based Cooper pair splitter with a Nb superconducting electrode. The large critical magnetic fields of Nb allow us to explore CPS up to 
$\sim1\,$T. We find that the conductance 
correlations cannot only manifest as symmetric peaks and dips in the gate dependence, 
but also as strongly asymmetric shapes reminiscent of Fano resonances. We interpret the experimental results 
in a minimal model that incorporates the superconducting 
proximity effect, the tunnel coupling between the QDs and quantum interference (see Fig.~1b). Interference results in asymmetric Fano-type features 
for local transport processes, whereas CPS produces Lorentzian shaped contributions to the conductances. These effects can be strong enough to obscure the positive correlations due to CPS.
The interference features in our experiments suggest that electrons 
can be transmitted coherently through a Cooper pair splitter device, which is a
fundamental prerequisite for testing Bell's inequality in a beam mixer 
setup \cite{Burkard_Loss_PRB61_2000}. Since often the spin-coherence 
time is longer than the spatial coherence, our results are also 
encouraging for the various propositions to detect spin entanglement
\cite{Kawabata_JPhysSocJpn70_2001, Braunecker_Levi_Yeyati_PRL111_2013, Klobus_Baumgartner_Martinek_PRB89_2014, Cottet_PRB86_2012, Scherubl_Csonka_PRB89_2014}.

Figure~\ref{Figure1}a shows a schematic of a CPS device. 
Using electron-beam lithography, an array of twelve $40\,$nm wide
local bottom gates ($4/18\,$nm Ti/Pt, $\sim 60\,$nm spacing) were fabricated on a
silicon substrate, electrically insulated by $\sim25\,$nm SiN$_x$.
A single NW ($\sim 70\,$nm diameter) was then deposited perpendicular 
to the gates using micromanipulators. The NWs were grown by 
solid-source molecular beam epitaxy
\cite{Madsen_Nygard_JCrystGrowth_2013}, using an optimized
process to suppress stacking faults \cite{Shtrikman_Heiblum_Nanolett_2009}. 
A $330\,$nm wide and $110\,$nm thick superconducting Nb contact in the center 
and two normal metal electrodes ($7$/$95\,$nm Ti/Au) at the ends of the NW were 
fabricated consecutively by conventional electron 
beam lithography, using an ammonium sulfide 
passivation \cite{Suyatin2007} to remove the native NW oxide.

The experiments were carried out in a dilution refrigerator at 
a base temperature of $\sim 50\,$mK. The QDs on either side of S were each induced in the NW by a negative voltage applied to 2 neighboring bottom gates, of which one was also used to tune the chemical potential of the respective QD. We label the voltages on these two tuning gates as $V_{\rm g1}$ and V$_{\rm g2}$ and use the terms 'local gate' and 'far gate' to distinguish the two gates when discussing a specific QD. A sinusoidal voltage 
$V^{\rm (ac)}\approx 10\,\mu$V applied to S 
results in the simultaneously recorded currents $I_1^{(\rm ac)}$ 
and $I_2^{(\rm ac)}$ in the contacts N1 and N2, which were held at carefully nulled potentials. We define the differential 
conductance through each QD as $G_i=I_i^{\rm (ac)}/V^{\rm (ac)}$, both of which show well-defined, uncorrelated Coulomb blockade diamonds \cite{supplementary}.

\begin{figure}[t]
\centering
\includegraphics[width=3.in,clip]{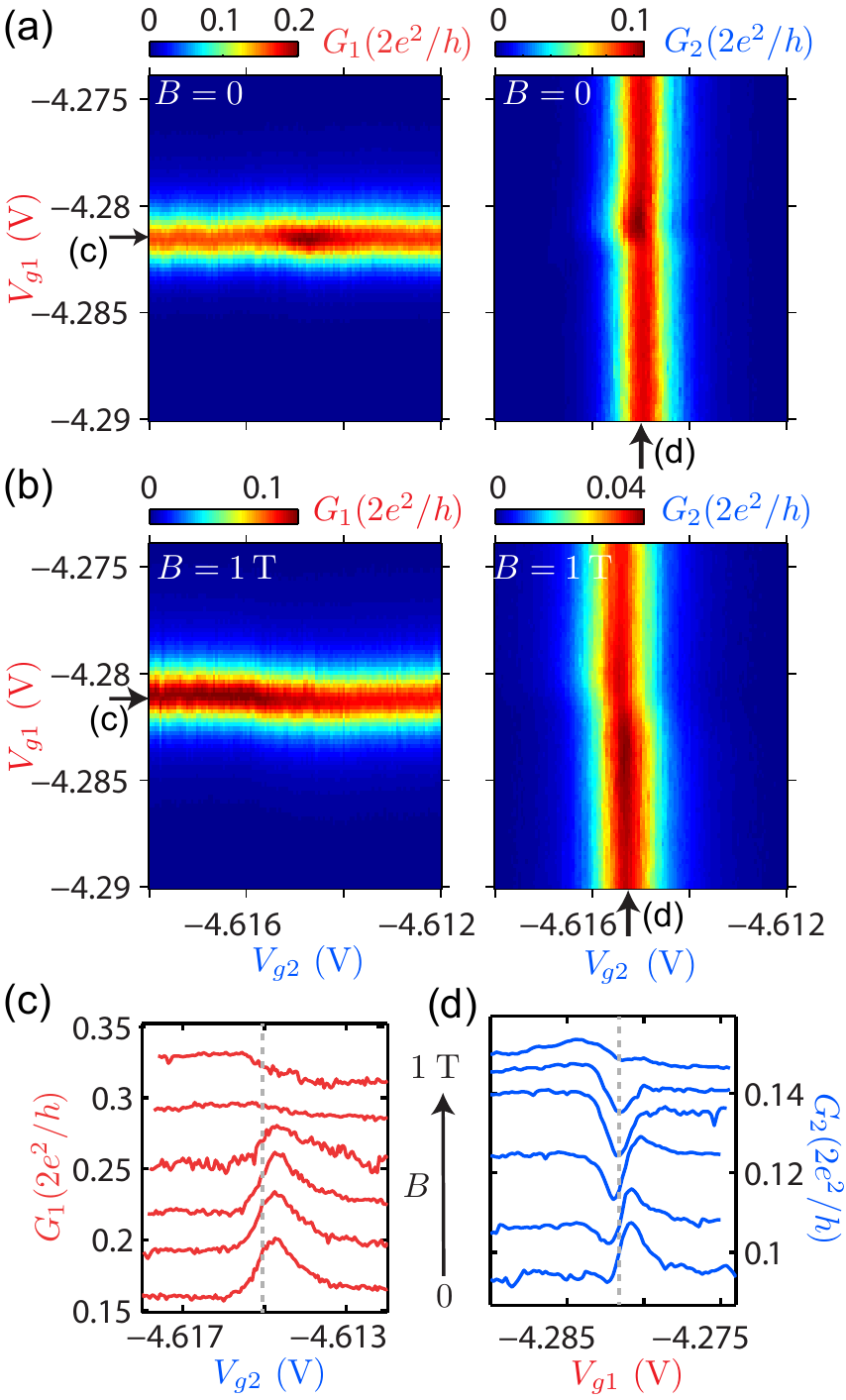}
\caption{(Color online) (a) and (b) $G_1$ and $G_2$ as a function of V$_{\rm g1}$ and V$_{\rm g2}$ for $B=0$ and $B=1\,$T, respectively.
(c) and (d) $G_1$ and $G_2$ as a function of the respective far gate,
V$_{\rm g2}$ and V$_{\rm g1}$, for a series of
magnetic fields $B$ with the local gates set to a QD resonance, as indicated by black arrows in Fig.~a and b. The curves are
shifted vertically for clarity.
}
\label{Figure2}
\end{figure}

Figures~2a shows $G_1$ and $G_2$ as a function of the two QD tuning gates for $B=0$. The resonance amplitudes of both QDs are increased when both are on resonance (resonance crossing), which results in a positive correlation between the conductance variations. In contrast, in the measurements at $B=1\,$T in Fig.~2b, the conductance maxima do not occur exactly at the expected resonance crossings, but vary in position, which results in a conductance not symmetric with respect to a crossing. Figs.~2c and d show cross sections taken on top of the a QD resonance as a function of the respective far gate for a series of magnetic fields. Away from a resonance crossing the resonance positions and amplitudes are independent of the respective far gate, while near a crossing both vary with gate voltage and exhibit different symmetries with respect to the resonance crossing (dashed lines) with increasing field.
For $B=0$ both resonance modulations are dominated by a roughly Lorentzian 
shape with a small minimum at the low-voltage side of the resonance crossing. As we increase the magnetic field, $G_1$ and $G_2$ evolve differently: while $G_1$ changes only little up to $B \approx 400\,$mT and then becomes broadened at higher fields, the variation of $G_2$ first evolves into a dip/peak structure, into a dip at intermediate fields and a maximum at the low-voltage side of the resonance crossing for the highest fields. The finite-field curves are reminiscent of Fano-type resonances, rather than Lorentzians.
Directly comparing the $B=0$ to the $B=1\,$T curve, one finds that the conductance maximum is shifted from the right to the left side of a crossing. We note that for most curves the maxima and minima in the conductance variations do not coincide with the resonance crossing.

The resemblance of the experimental conductance variations to Fano-type resonances 
suggests that at least one of the transport mechanism is prone to quantum interference, which cannot be described in the incoherent models of previous works 
\cite{Fulop_dHollosy_Baumgartner_PRB90_2014, Schindele_Baumgartner_PRB89_2014}. 
To account for the observed interference patterns also requires to go beyond earlier coherent two-dot models \cite{Herrmann_Kontos_Strunk_PRL104_2010, Chevallier_Martin_PRB82_2010, Eldridge_Koenig_PRB82_2010}, where the QDs are coupled by direct tunneling. Here we introduce a 3-site model, shown schematically in Fig.~\ref{Figure1}b, in which each QD is represented 
by a single spin-degenerate level with Coulomb interactions and coupled coherently to a central NW segment below S and to the leads N1 and N2. The central segment is also modeled as a single level coherently coupled to S.
The Hamiltonian (without the coupling to the leads \cite{supplementary}) reads
\begin{align}
H_{0}=\sum_{\sigma,i} \epsilon_{i,\sigma} n_{i,\sigma} + U_i n_{i,\uparrow}n_{i,\downarrow}+ \sum_{\sigma, i\neq m}t_{im} d^\dagger_{i,\sigma}d_{m,\sigma}+{\text h.c.}, \nonumber
\end{align}
where $i=1,m,2$ label the left, middle and right levels with
the on-site energies $\epsilon_{i,\sigma}=\epsilon_i(B)+\sigma g_i B/2$, comprising orbital and Zeeman shifts. For the g-factors $g_i$ we use typical values in the range of 
$\left|g\right|=5-15$ \cite{Csonka_Hofstetter_NanoLetters_2008, dHollosy_Baumgartner_AIPProc_2012}. $t_{im}$ 
corresponds to the hopping amplitudes from the QDs to the central site.
To simulate the superconducting proximity effect we set the Coulomb interactions on the central region to zero, $U_m=0$, which 
is justified by the large size and the screening by the superconductor. 
This assumption allows the occupation of the central level by two electrons, e.g. a Cooper pair from S. We assume that only this central region is coupled to S.
$\gamma_i$, with $i=1,m,2$, are the tunneling rates from the three sites to the respective lead, see Fig.~1b.
We calculate the linear transport characteristics using an equation of motion approach for the electronic Green's functions 
\cite{Burset2011a, supplementary}. 
The conductances can be decomposed as
$G_{i} = G_{\rm{loc};i} + G_{\rm{CPS}}$ 
where  the local contributions comprise local Andreev reflection (LAR) and single quasiparticle tunneling, $G_{\rm{loc};i}=G_{\rm{LAR};i} + G_{\rm {qp};i}$. $G_{\rm{CPS}}$ denotes the contribution due to CPS.

\begin{figure}[t]
\centering
\includegraphics[width=3.in,clip]{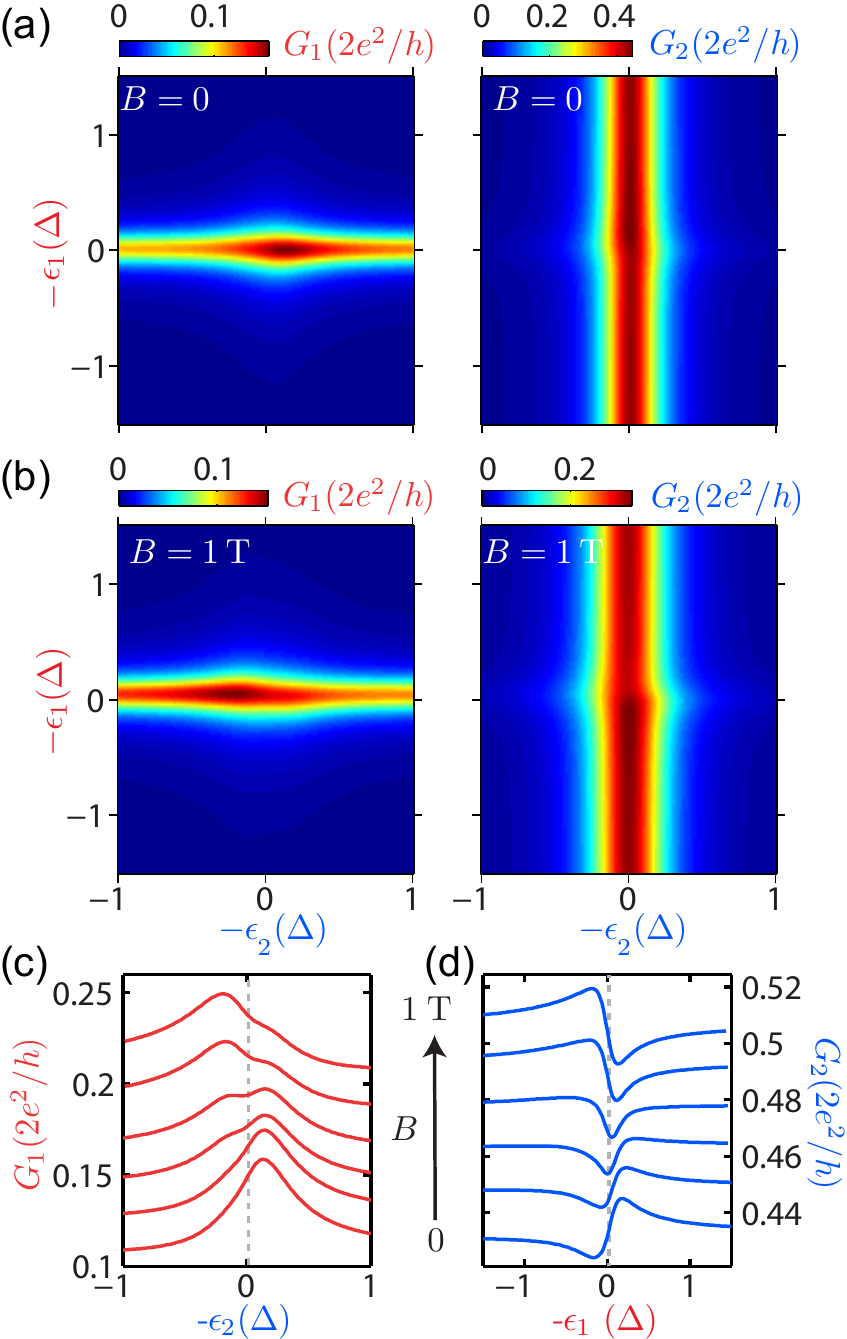}
\caption{(Color online) Conductances through QD1 and QD2 
in the three-site model with parameters chosen to qualitatively 
reproduce the experiments in Fig.~\ref{Figure2}. (a) and (b) $G_1$ and $G_2$ as a function of both QD level positions. (c) and (d) On-resonance conductance variations of QD1 and QD2, respectively, as a function of the level position of the other QD, for increasing values of magnetic field. The parameters used in the calculations are 
$\epsilon_m=-0.03\Delta$, $2t_{1m}=t_{2m}=0.1\Delta$, U=3$\Delta$, 
and the coupling to the leads $\gamma_1=0.15\Delta$, $\gamma_2=0.2\Delta$, and $\gamma_m=0.11\Delta$.
}
\label{Figure3}
\end{figure}

The 3-site model qualitatively reproduces the experiments in Fig.~2. The corresponding calculated transport characteristics are plotted in Fig.~\ref{Figure3} for typical experimental parameters. Here, $-\epsilon_i$ corresponds to the changes in gate voltage (up to a lever arm factor). The model accounts for 
the shifts of the resonance maxima and minima with
increasing magnetic field, the relatively weak change in $G_1$ at low fields, as well as the peak/dip/peak transitions in $G_2$ with increasing field.

To intuitively understand the structure of the conductance variations and the field evolution, we plot in Fig.~\ref{Figure4}a the local and CPS contributions to the total conductance $G_2$ for $B=0$ and $B=1.2\,$T. 
We find that $G_{\rm loc}$ of both QDs often exhibits a strong asymmetric dip as a function of the respective far gate voltage, 
while $G_{\rm CPS}$ generally results in a Lorentzian peak, not necessarily at the same position.
The physical reason for these characteristics is the impact of interference:
for the local processes into N2, the lowest order non-zero contribution 
comprises two different paths. Either an electron tunnels directly from the central region to QD2 and to lead N2, or it reaches QD2 and N2 after an excursion to QD1 and tunneling through the middle region, see Fig.~\ref{Figure1}b. 
Since the transmission phase acquired on a QD state can vary by a large fraction of $\pi$ \cite{Schuster_Shtrikman_Nature385_1997}, the two paths can interfere constructively or destructively, depending on the relative level positions, resulting in a minimum or a maximum in the local contribution and in an asymmetric conductance variation.
In contrast to the local currents, the lowest-order non-zero CPS contribution stems from the direct tunneling of one electron into each QD, which is
not affected by interference \cite{supplementary}. The sum of the local and nonlocal processes can then result in the observed Fano-type line shapes.
 
The inversion of the line shape symmetry with magnetic field, shown in Fig.~4a, can be understood as follows:
the orientation of a Fano resonance depends on a generic shape-parameter \cite{Fano1961}, which is given here essentially by $\epsilon_m$ \cite{supplementary}. When $\epsilon_m$ is below (above) the Fermi energy, the asymmetry of the local contribution has a maximum at a more positive (negative) gate voltage than the dip.
In the experiments of Figs.~\ref{Figure2}c and d we find a transition from a peak/dip to a dip/peak structure at $B\approx 0.6\,$T, which we thus identify as the field at which the position 
of the central level is shifted from below to above the Fermi energy by orbital and Zeeman shifts. We note that an asymmetric shape also shifts the maxima of the total conductance with respect to the resonance crossing. We observe these characteristics in the experiments (Figs.~2c and d) as well as in our calculations (Figs.~3c and d), where the maxima and minima do not occur at the dashed lines that indicate the crossings.

As can be inferred from Fig.~4a, the CPS contribution does not change significantly with $B$ (for low enough $B$), but the local contribution can be reduced strongly. The latter is due to the lifting of the spin-degeneracy by the Zeeman shift, which results in a reduction of LAR due to the opposite-spin electrons of a Cooper pair tunneling through a (partially) polarized level. This leads to an increase in the CPS efficiency defined as $\chi=2G_{\rm CPS}/(G_1+G_2)$ \cite{Hofstetter2009, Schindele_Baumgartner_Schoenenberger_PRL109_2012}.

\begin{figure}[t]
\centering
\includegraphics[width=3.in,clip]{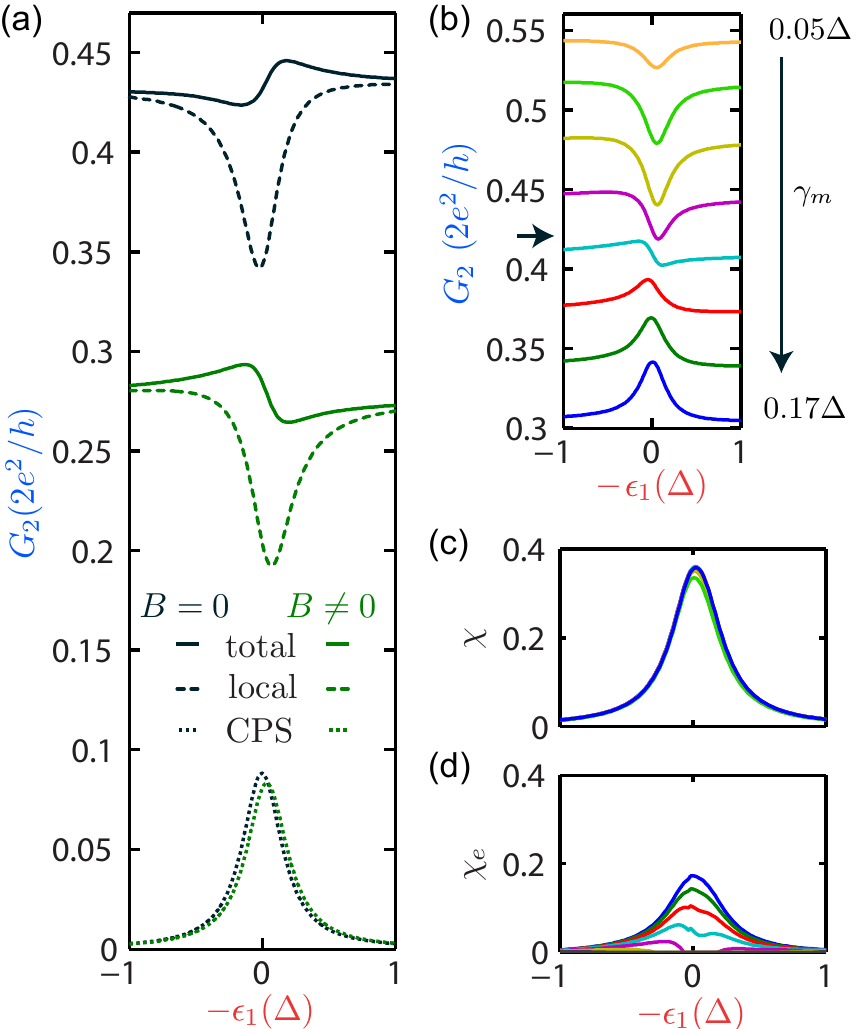}
\caption{(Color online) (a) Total conductance
through the QD2, $G_2$ (solid line), and the decomposition 
into the local (dashed line) and CPS (dotted line) contributions, as a function of $-\epsilon_1$ for $B=0$ (dark) and and $B=1.2\,$T (green).
(b) $G_2$ vs. V$_{g1}$ for 
different coupling strengths $\gamma_m$ to the superconductor.
The curves are offset vertically for clarity.
The value at which $\gamma_m=t_{im}$ is pointed out by an arrow.
(c) and (d) Correct and estimated CPS efficiencies, $\chi$ and $\chi_{\rm e}$, for the numerical results in (b).}
\label{Figure4}
\end{figure}

The efficiency is relevant for most prospective applications of a CPS device \cite{Schindele_Baumgartner_Schoenenberger_PRL109_2012, Klobus_Baumgartner_Martinek_PRB89_2014}. In the above expression for $\chi$ one usually estimates $G_{\rm CPS}$ by subtracting the conductance sufficiently far away from the resonance crossing. We call this experimentally obtained efficiency $\chi_{\rm e}$. It was shown for an incoherent model that this procedure underestimates the actual CPS efficiency \cite{Fulop_dHollosy_Baumgartner_PRB90_2014}. However, this is not obvious for coherent transport. In Fig.~4a we show in an example that the local contribution is not constant, but exhibits a minimum at the resonance crossing, which also suggests that $\chi\geq\chi_{\rm e}$.
To further illustrate this effect, we have calculated $G_2$ vs. $\epsilon_1$ (i.e. $V_{g1}$) in the 3-site model for a series of tunnel coupling strengths $\gamma_m$ between S and the middle level, see Fig.~4b. We find that the conductance minimum for weak couplings evolves into a maximum for $\gamma_m \gtrsim t_{im}$ (black arrow). The reason for this transition is an additional contribution by states forming in the superconducting gap as $\gamma_m$ increases.
In Fig.~4c, $\chi$ is plotted for all curves in Fig.~4b and is almost identical for all $\gamma_m$. For comparison, $\chi_{\rm e}$ is plotted in Fig.~4d for the same conductances, which demonstrates that $\chi_{\rm e}$ systematically underestimates $\chi$ (we have omitted negative values) \cite{footnote}.

In summary, we report experiments on a Cooper pair splitting device that exhibits characteristics of quantum interference, which are tunable by electrical gates and external magnetic fields. To interpret these data, we introduce a 3-site model that reproduces the experiments on a qualitative level and allows the decomposition of the conductances
into local and CPS contributions. Specifically, we show that quantum interference is relevant in such structures, which is required for any application of CPS as a source of entangled electron pairs. In addition, we discuss the impact of various device parameters and find that the experimentally estimated CPS efficiency is quite generally a lower bound for the actual value also in the coherent transport regime. The presented model might also be used to gain a new vantage point for other fundamental effects in superconducting proximity systems, such as fractional fermions or Majorana 
bound states. For example, a similar setup could be used to detect the inversion of the gap in a Majorana wire by the change of symmetry in the local Andreev conductance as a single level crosses the Fermi energy.

We gratefully acknowledge the financial support by the EU FP7 project SE$^2$ND, the EU ERC projects CooPairEnt and QUEST, SCIEX project NoCoNano, 
the Swiss NCCR Quantum, the Swiss SNF, and the Danish Research Councils.

\bibliographystyle{apsrev}

\begin{thebibliography}{99}
\bibitem{footnote}
\bibinfo{note}{Clearly, $\chi_e$ overestimates $\chi$ for $\epsilon_m \rightarrow 0$ where the CPS contribution is masked by the global conductance dip, but we note that this also holds for $\epsilon_m >> t_{im}$ due to the misalignment of the peaks in the local contribution.}
\end{thebibliography}


\clearpage
\onecolumngrid

\newcommand{\void}[1]{}
\newcommand{\comment}[1]{\textit{$\blacktriangleright$~#1~$\blacktriangleleft$}}
\renewcommand{\widehat}{}
\newcommand{\tr}{\mathop{\mathrm{Tr}}\nolimits}

\renewcommand\thefigure{S\arabic{figure}} 

\section{Supplementary Material}

\section{Quantum dot characterization}

The Cooper pair splitter device was equipped with 12 bottom gates 
(see Fig.~1a in the main text). Starting the numbering on the left, 
we applied negative voltages to the 2nd and 3rd gate to form the barriers for a quantum dot (QD) in the left arm (QD1), and to the 9th and 10th gate to form QD2 in the right arm. Gate 2 was then used to tune QD1 and labeled ${g1}$ in the main text. Similarly, gate 8 was used to tune QD2 and labeled ${g2}$. Representative conductance maps of QD1 and QD2 as a 
function of the bias applied to S and the respective tuning gate voltage are presented in Fig.~S1.

\begin{figure}[h]
\centering
\includegraphics[width=3.in,clip]{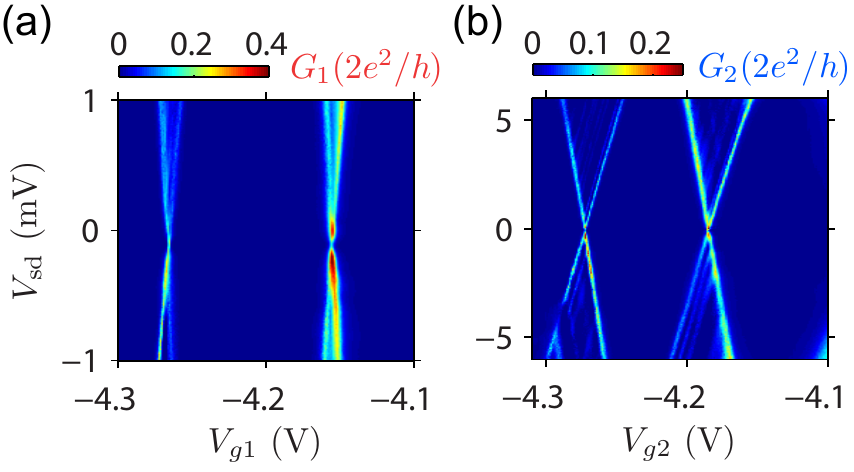}
\caption{(Color online) (a) and (b) Representative conductance maps of 
QD1 and QD2, respectively, as the 
function of the bias and the respective tuning gate voltage. 
The charging energies are in the range of $E_{\rm c} \approx 2-3\,$meV for 
QD1, and $E_{\rm c} \approx 8-12\,$meV for QD2. 
The FWHM (full-widths-half-maximum) are $\sim 130-200\,\mu$eV for both QDs.
\label{Figure4}}
\end{figure}

\section{Model Hamiltonian for interacting electrons: self-consistent approach}

In this section we provide a description of the self-consistent 
method used for the calculations in the main text \cite{Burset}.

{\it Model and Hamiltonian}---
We start by defining the model Hamiltonian for the three tunnel-coupled sites in our 3-site model (see Fig.~1b of the main text),
\begin{align}
H_{0}=\sum_{\sigma,i} \epsilon_{i,\sigma} n_{i,\sigma} + U_i n_{i,\uparrow}n_{i,\downarrow}+ \sum_{\sigma, i\neq m}t_{im} d^\dagger_{i,\sigma}d_{m,\sigma}+{\text h.c.}.
\label{eq-ham}
\end{align}

As described in the main text, QD1 and QD2 are coupled to the normal leads 1 and 2, respectively, each described as non-interacting Fermi
liquids with a Hamiltonian of the form
\begin{align}
H_{l}=\sum_{{\bf k}\sigma}\epsilon_{\bf k}a_{l{\bf k}\sigma}^{\dagger}a_{l{\bf k}\sigma}.
\end{align}
$a_{l{\bf k}\sigma}^{(\dagger)}$ are the annihilation (creation) operators in for electrons in lead {\it l} with momentum {\bf k} and spin $\sigma$.
$t_{im}$ is the tunneling amplitude between the central site and site {\it i}. 

The central site is coupled to the superconducting lead S, for which we use the
BCS-Hamiltonian
\begin{align}
H_{S}=\sum_{{\bf k},\sigma} \xi_k c_{{\bf k}\sigma}^\dagger c_{{\bf k}\sigma} -\sum_k (\Delta_{\bf k} c_{{\bf k}\uparrow}^\dagger c_{-{\bf k}\downarrow}^\dagger+
\Delta_{\bf k}^* c_{-{\bf k}\downarrow} c_{{\bf k}\uparrow})
\end{align}
where $c_{{\bf k},\sigma}^\dagger$ creates a fermion with momentum {\bf k} and spin $\sigma$. 

The coupling to the leads is given by
\begin{align}
H_\tau = 
\sum_{\substack{{\bf k},\sigma \\ i=1,2}} t_i \left( d_{i,\sigma}^{\dagger} a_{i,{\bf k},\sigma} + 
\mathrm{h.c.}\right)  + 
\sum_{{\bf k},\sigma } t_{s} \left(d_{m\sigma}^{\dagger} c_{{\bf k}\sigma} + \mathrm{h.c.}\right)
\label{eq.coupling}
\end{align}
Tunneling from a QD to the state ${\bf k}$ 
in the normal lead $l$ is described by the amplitude $t_{l}$.
We assume that the ${\bf k}$-dependence of the amplitudes $t_{l}$ and $t_s$
can be neglected (wide band limit). Thus, we can define the tunnel rates between site {\it i} and 
lead {\it i} as $\gamma_i=\pi \rho_i t_i^2$ (for {\it i}=1,~m~and~2), 
with $\rho_i$ the density of states per spin in lead {\it i}.

{\it Green's function of the interacting system}--- 
In the presence of the Coulomb interaction, it is not possible to obtain the exact Green's function (GF). We therefore decouple the corresponding equation of motion (EOM) at the level of two-body GFs. Under this approximation the one-body GF
\begin{align}
G(t,t')={\it i}\Theta(t'-t) \langle\langle \{ \psi(t), \psi(t')^\dagger \} \rangle\rangle
\label{g1}
\end{align}
is coupled to the two-body GF by \cite{Burset} 
\begin{align}
W(t,t')={\it i}\Theta(t'-t) \langle\langle \{ \xi(t), \psi(t')^\dagger \} \rangle\rangle,
\label{g2}
\end{align}
which can be written in the basis
\begin{align}
&\psi^\dagger(t)=\begin{bmatrix}
d_{1\uparrow}^\dagger & d_{1\downarrow} & d_{m\uparrow}^\dagger &  d_{m\downarrow}   & d_{2\uparrow}^\dagger   & d_{2\downarrow} 
\end{bmatrix}_t
\\
&\xi^\dagger(t)=\begin{bmatrix}
d_{1\uparrow}^\dagger n_{1,\downarrow}& d_{1\downarrow}n_{1,\uparrow} & d_{m\uparrow}^\dagger n_{m,\downarrow} &  d_{m\downarrow} n_{m,\uparrow}   & d_{2\uparrow}^\dagger n_{2,\downarrow}  & d_{2\downarrow} n_{2,\uparrow} 
\end{bmatrix}_t.
\end{align}
The dot operators are used in the Heisenberg picture labeled as `t'. The 3-sites model leads to a $12\times 12$ GF, but
in the absence of a magnetic field we can use the time reversal symmetry 
to express the Green's function in a reduced $6\times 6$ matrix form. 

The time derivative of Eqs.~\eqref{g1} and~\eqref{g2} yields the equation of motion (EOM) with the Fourier transform
\begin{align}
&(\hat{\omega}-\hat{h}_0 -\hat{\Sigma} - \hat{\Gamma}) G (\omega)={\bf 1}  + \hat{U} W(\omega),~\text{and}
\label{GW1}
\\
& W(\omega)=(\omega-\hat{h}_0-\hat{U})^{-1} \hat{A} \left[ {\bf 1} +\hat{\Sigma}+\hat{\Gamma} \right] G(\omega),
\label{GW2}
\end{align}
where we omit terms involving Kondo-like correlations.
We have used the matrices $\hat{h}_0=\sum_{i,j=l,m,r} (-1)^{i+1}\delta_{i,j}\epsilon_i$, 
\begin{equation}
\hat{\Sigma}= \begin{pmatrix}
                  0     &     0      &    t_{1m}            &   0             &    0     & 0 \\
                  0     &     0      &      0               &  -t_{1m}        &    0     & 0 \\
		  t_{1m}&     0      &      0               &   0             &    t_{2m}& 0 \\
		  0     &     -t_{1m}&      0               &   0             &    0     & -t_{2m}\\
		  0     &     0      &   t_{2m}             &   0             &    0     & 0 \\
		  0     &     0      &      0               &  -t_{2m}        &    0      & 0  \\
          \end{pmatrix},~~~~
\hat{U}= \begin{pmatrix}
                  U     &     0      &      0               &   0             &    0     & 0  \\
                  0     &     -U     &      0               &   0             &    0     & 0  \\
		  0     &     0      &      0               &   0             &    0     & 0  \\
		  0     &     0      &      0               &   0             &    0     & 0  \\
		  0     &     0      &      0               &   0             &    U     & 0  \\
		  0     &     0      &      0               &   0             &    0     & -U  \\
          \end{pmatrix}
\end{equation}
\begin{align}
\hat{A}= 
\begin{pmatrix}
n_{1,\downarrow}     &      \langle d_{1,\downarrow} d_{1,\uparrow}\rangle      &  0    &  0    &0     &0 
\\
\langle d_{1,\downarrow}^\dagger d_{1,\uparrow}^\dagger\rangle     &       n_{1,\uparrow}      &  0      & 0     & 0   &0
\\
0   &     0    &   n_{m,\downarrow} &   \langle d_{m,\downarrow} d_{m,\uparrow}\rangle   &    0    & 0 
\\
0   &  0    &  \langle d_{m,\downarrow}^\dagger d_{m,\uparrow}^\dagger\rangle      &   n_{m,\uparrow}    &   0    & 0
\\
0 &0  &0       & 0           &    n_{2,\downarrow}    &  \langle d_{2,\downarrow} d_{2,\uparrow}\rangle   \\
0    &     0     &  0           &  0     &   \langle d_{2,\downarrow}^\dagger d_{2,\uparrow}^\dagger\rangle     & n_{2,\uparrow} \\
\end{pmatrix}
\end{align}
and
\begin{align}
\hat{\Gamma}=
\begin{pmatrix}
                  {\it i}\gamma_1     &   0     &   0           &   0             &    0     & 0 \\
                 0   &     {\it i}\gamma_1     &      0             &   0           &    0     & 0 \\
		  0  &     0      &      g_m(\omega)          &  -\Delta_{ind}(\omega) & 0   & 0 \\
		  0     &     0    &   -\Delta_{ind}(\omega)    &   g_m(\omega)       &   0      & 0\\
		  0     &     0      &   0             &   0             &  {\it i}\gamma_2    & 0\\
		  0     &     0      &      0             &  0      &   0      & {\it i}\gamma_2 \\
          \end{pmatrix}.
\end{align}
where the coupling to the normal contacts gives rise to the self-energies $\gamma_i$, and 
the coupling to the superconductor results in the self energies

\begin{align}
&g_m(\omega)=-\gamma_{m}\frac{\omega-{\it i}\eta_s}{\sqrt{\Delta^2 -(\omega-{\it i} \eta_s)^2}}\\
&\Delta_{ind}(\omega)=\gamma_{m}\frac{\Delta}{\sqrt{\Delta^2 -(\omega-{\it i} \eta_s)^2}}.
\end{align}
The first term represents the self energy due to the quasiparticles in the superconductor, while the second term is the anomalous self energy that induces a superconducting gap on the central site.

Substituting Eq.~\eqref{GW2} into Eq.~\eqref{GW1}, we obtain
\begin{align}
G (\omega)=\left(g_0^{-1}-\hat{\Sigma}-\hat{\Gamma}\right)^{-1}
\label{final}
\end{align}
where 
\begin{align}
g_0=(\omega-\hat{h}_0)^{-1}\left({\bf 1} + \hat{U} (\omega-\hat{h}_0-\hat{U})^{-1} \hat{A}\right).
\end{align}
Equation~\eqref{final} needs to be calculated self-consistently because 
the matrix $\hat{A}$ contains parameters ($ n_{i,\sigma}$ and $\langle d_{i,\downarrow} d_{i,\uparrow}\rangle$) 
that depend on the resulting $G(\omega)$.

{\it Linear conductance}--- In the linear response regime, with symmetric biasing and $T\rightarrow 0$, 
the conductance through QD {\it l} contains four contributions that can be expressed in terms of the 
advanced Green's functions of the system, 
\begin{align}
&G_{CPS}=2\gamma_1\gamma_2 |G_{e_1,h_2}|^2\\
&G_{LAR;l}=4\gamma_l^2 |G_{e_l,h_l}|^2\\
&G_{qp;l}= 2\gamma_l {\rm Im}\{ g_m(0)\} |G_{e_l,e_m}|^2 \\
&G_{EC}=0
\end{align}
where the arguments of the Green's function $e_i$ and $h_i$ refer to the 
electron and hole positions of the site {\it i}.
The different contributions to the conductance are Cooper pair splitting, $G_{CPS}$, local Andreev and the quasiparticle processes through QD $l$,
$G_{LAR;l}$ and $G_{qp;l}$, and elastic cotunneling, $G_{EC}$. 

\section{Non-interacting case}

In this section, 
we derive an analytical expression for the total conductance through a Cooper pair splitter for the non-interacting Hamiltonian (with $U_i=0$ in Eq.~\eqref{eq-ham}).
Using standard techniques \cite{Datta} we construct 
the exact advanced Green's function from the 
Hamiltonians introduced in the previous section. 
For this purpose it is convenient to define the Nambu basis
\begin{equation}
\psi^\dagger(t)=\begin{bmatrix}
d_{1\uparrow}^\dagger& d_{m\uparrow}^\dagger & d_{2\uparrow}^\dagger &  d_{1\downarrow}   & d_{m\downarrow}   & d_{2\downarrow} 
\end{bmatrix}_t,
\end{equation}
where `t' stands for the Heisenberg picture used for the dot operators.
Then, we obtain the Green function 
\begin{equation}
G(\omega)= \begin{pmatrix}
                  \omega-\epsilon_1-{\it i}\gamma_1     &     -t_{1m}      &    0              &   0             &    0     & 0 \\
                  -t_{1m}     &     \omega-\epsilon_m-g_m(\omega)   &     -t_{2m}               & 0        &    \Delta_{ind}(\omega)     &            0 \\
		     0         &       - t_{2m}    &      \omega-\epsilon_2-{\it i}\gamma_2     &     0             &    0&     0 \\
		     0         &     0           &      0      &   \omega+\epsilon_1-{\it i}\gamma_1             &    t_{1m}     & 0\\
		     0         &     \Delta_{ind}(\omega)      &      0      &   t_{1m}             &    \omega+\epsilon_m-g_m(\omega)     & t_{2m} \\
		     0         &          0      &      0      &   0             &    t_{2m}      & \omega+\epsilon_2-{\it i}\gamma_2  \\
          \end{pmatrix}^{-1},
          \label{eq-GF}
\end{equation}
One can obtain analytical expressions of the conductances from Eq.~\eqref{eq-GF}. 
However, these expressions are cumbersome and it is more instructive to expand Eq.~\eqref{eq-GF} perturbatively in terms of the induced gap 
$\Delta_{ind}$, that is, $\epsilon_m,~t_{im} \gg \Delta_{ind}$. 
Thus, to first order in $\Delta_{ind}$ we obtain
\begin{equation}
G= \begin{pmatrix}
                  \hat{\omega}+H_{e}            &     \hat{\Delta}    \\
                  \hat{\Delta}     &     \hat{\omega}+H_{h}\\
          \end{pmatrix}^{-1}\approx 
          \begin{pmatrix}
                  (\hat{\omega}+H_{e})^{-1}            &    - (\hat{\omega}+H_{e})^{-1}\hat{\Delta}(\hat{\omega}+H_{h})^{-1}    \\
                  -(\hat{\omega}+H_{h})^{-1}\hat{\Delta}(\hat{\omega}+H_{e})^{-1}      &    (\hat{\omega}+ H_{h})^{-1}\\
          \end{pmatrix},
\end{equation}
where we defined the matrix $\hat{\omega}=\omega \sum_{i,j=1}^3 \delta_{i,j}$, and
\begin{equation}
H_{e/h}= \begin{pmatrix}
                  \mp \epsilon_1-{\it i}\gamma_1     &     \mp t_{1m}      &    0   \\
                  \mp t_{1m}     &     \mp \epsilon_m-g_m(\omega)   &     \mp t_{2m}              \\
		     0         &        \mp t_{2m}    &     \mp \epsilon_2-{\it i}\gamma_2      \\
          \end{pmatrix}~~~\text{and}~~~~
          \hat{\Delta} = \begin{pmatrix}
                  0     &    0             &  0  \\
                  0     &   \Delta_{ind}   &  0   \\
		  0     &        0         &  0    \\
          \end{pmatrix}. \nonumber
\end{equation}

Then, we obtain the corresponding conductances
\begin{align}
&G_{CPS}\sim 2\gamma_1\gamma_2\frac{  t_{1m} t_{2m}\gamma_m^2}{\epsilon_m^2(\epsilon_1^2+\gamma_1^2)(\epsilon_2^2+\gamma_2^2)}, \\
&G_{LAR;l}\sim 4\gamma_l^2 \frac{t_{lm}^4 \gamma_{m}^2}{\epsilon_m^4(\epsilon_l^2+\gamma_l^2)^2 }\left(1+\frac{2\epsilon_j  t_{jm}^2}{\epsilon_m(\epsilon_j^2+\gamma_j^2)}\right)^2~~~\text{and}\\
\label{eqlar}
&G_{qp;l}\sim \frac{2 t_{lm}^2 \gamma_l \text{Im}\{ g_m(0)\}}{\epsilon_m^2(\epsilon_l^2+\gamma_l^2)}  \left(1+\frac{\epsilon_j  t_{jm}^2}{\epsilon_m(\epsilon_j^2+\gamma_j^2)}\right)^2.
\end{align}
The {\it j}-index stands for the opposite dot of QD {\it l}.
From these expressions we find that CPS has a Lorentzian and the 
LAR and qp contributions a Fan-type shape. 
From the LAR and qp expressions one finds that the asymmetry orientation depends on the sign of $\epsilon_m$, which one can thus relate to the shape-parameter in a Fano-type resonance \cite{Fano1961}. 
We note that $G_{LAR;l}$ and $G_{qp;l}$ exhibit 
the same gate dependence except for a factor 2 in the interference term. This factor accounts for the two particles in the local Andreev reflection.
This factor is the origin of the more 
asymmetric interference patterns in the LAR contribution of the deeper minima compared to the quasiparticle contribution.

\end{document}